\documentstyle[12pt]{article}

\font\tenrm=cmr10

\begin{document}

\newcommand{\beq}{\begin{equation}}
\newcommand{\eeq}{\end{equation}}
\newcommand{\beqa}{\begin{eqnarray}}
\newcommand{\eeqa}{\end{eqnarray}}
\newcommand{\beqar}{\begin{eqnarray*}}
\newcommand{\eeqar}{\end{eqnarray*}}
\newcommand{\tr}{{\rm tr}}
\newcommand{\be}{\beta}
\newcommand{\al}{\alpha}
\newcommand{\ie}{{\it i.e.,}\ }
\newcommand{\eg}{{\it e.g.,}\ }
\newcommand{\ch}{{\cal H}}
\newcommand{\ssc}{\scriptscriptstyle}
\def\d{\delta}
\def\eps{\epsilon}
\def\heps{\hat{\epsilon}}
\def\beps{\bar{\epsilon}}
\def\rL{\widetilde{L}}
\def\tc{\tilde{\chi}^a}
\def\tchi{\tilde{\chi}}
\def\ddx{d^4\!x}
\def\dtx{d^{2}\!x}
\def\S{{\cal S}}

\renewenvironment{thebibliography}[1]
  { \begin{list}{\arabic{enumi}.}
    {\usecounter{enumi} \setlength{\parsep}{0pt}
     \setlength{\itemsep}{3pt} \settowidth{\labelwidth}{#1.}
     \sloppy
    }}{\end{list}}

\parindent=1.5pc

 ~
\vskip -0.5truein
\rightline{\small gr--qc/9502009\hfill McGill/95--04; UMDGR--95-092}
\vskip 0.4truein


\begin{center}{{\bf BLACK HOLE ENTROPY}\\
{\bf IN HIGHER CURVATURE GRAVITY}\footnote{Based on talks
presented by RCM at the
16$^{\rm th}$ Annual Montr\'eal-Rochester-Syracuse-Toronto (MRST)
Meeting, {\sl What Next? Exploring the Future of High-Energy Physics},
held at McGill University, Montr\'eal, Canada, 11--13 May 1994, and at
{\sl Heat Kernel Techniques and Quantum Gravity} held at University of
Manitoba, Winnipeg, Canada, 2--6 August 1994.}\\
\vglue 1.0cm
{TED JACOBSON and GUNGWON KANG}\\
\baselineskip=14pt
{\it Department of Physics, University of Maryland}\\
\baselineskip=14pt
{\it College Park, MD 20742--4111 USA}
\vglue 0.5cm
{ROBERT C. MYERS}\\
\baselineskip=14pt
{\it Department of Physics, McGill University}\\
\baselineskip=14pt
{\it Montr\'eal, Qu\'ebec, Canada H3A 2T8}\\


\vglue 0.8cm
{\tenrm ABSTRACT}}
\end{center}
{\rightskip=3pc
 \leftskip=3pc
 \tenrm\baselineskip=12pt
 \noindent
We discuss some recent results on
black hole thermodynamics within the
context of effective gravitational actions including higher-curvature
interactions. Wald's derivation of the First Law
demonstrates that black hole entropy can always be expressed
as a local geometric density integrated over a space-like
cross-section of the horizon. In certain
cases, it can also be shown that these entropy
expressions satisfy a Second Law. One such simple example
is considered from the class of higher curvature
theories where the Lagrangian consists of a polynomial in the Ricci
scalar.

\vglue 0.8cm}
%
%
{\bf\noindent 1. Introduction}
\vglue 0.4cm
\baselineskip=14pt
Thermodynamics and statistical mechanics describe
systems with a large number of degrees of freedom
by following the evolution of a
few macroscopic parameters, rather than trying to understand
all of the details of the microphysics. One such parameter
is entropy. Within the context of thermodynamics, entropy is a
measure of the degradation of energy in processes. For a thermal
system undergoing an infinitesimal change,
the First Law of thermodynamics states
\begin{equation}
T\,\delta S=\delta U+\delta W
\label{firsttherm}
\end{equation}
where $T$, $S$, $U$ and $W$ are the temperature, the
entropy, the internal energy and the work done, respectively.
On the left hand side, $T\,\delta S$ indicates the amount of energy
which becomes unavailable for work in future processes, \ie the heat loss.
In the context of statistical mechanics, entropy has quite
a different significance. Here, it
is a measure of one's lack of detailed
information about the microphysical state of
a system. The First Law then specifies how this imprecision
changes in some process. In fact,
the Second Law of thermodynamics dictates that in any process
the total entropy will never decrease, \ie $\delta S_{tot}\ge0$.
This law is then a statement that one's uncertainty about
the microscopic physics increases as systems evolve.
It is perhaps a surprise that consistency of the Second
Law on very large scales requires recourse to general relativity\cite{LL}.
Regarding the entire universe as a single closed system, one would expect
that it should have relaxed into equilibrium in the maximum
entropy configuration, in contradiction with observations.
Such an equilibrium has not been reached because of the
cosmological expansion of the universe,
and hence there is no inconsistency when this effect of general
relativity is taken into account.

Another remarkable connection between thermodynamics and gravity
arises in black hole physics --- namely,
black holes carry an intrinsic entropy. This result
relies on the fundamental property that a black hole is a
region of spacetime which is inaccessible to observations, and
an essential role is played by the event horizon, the boundary
between the regions observable and unobservable from infinity.
Consider a box carrying some thermal system travelling through spacetime.
If the box interacts with other external systems,  one may expect
that its internal state will be taken out of equilibrium.
According to the Second Law, the subsequent evolution
would then be characterized by a continued increase of the entropy
as the system returns to equilibrium. If the box were
to fall into a black hole,
it would move out of the region
of spacetime in which measurements can be observed from infinity,
and there would no longer be any evidence of the entropy carried by
the box. The entropy in the observable spacetime would
thus appear to have decreased,
yielding an apparent violation of the Second Law.
To restore the validity of the Second Law, one can assign an extra
entropy to the boundary of the recorded spatial region (at each instant of
time). This boundary is, of course, a spatial cross-section of the
event horizon. Thus, by these very simple considerations, one is led
naturally to the concept of black hole (or more generally  horizon)
entropy.

Similar reasoning led
Bekenstein\cite{bek} to make the bold conjecture
some twenty years ago that, within general relativity, black
holes carry an intrinsic entropy given by the surface area of the
horizon measured in Planck units multiplied by
a dimensionless number of order one.
This conjecture was also suggested by Hawking's area
theorem\cite{areathm} which had shown that, like entropy, the horizon area
can never decrease in classical general relativity.
Bekenstein offered arguments for the proportionality of entropy and
area, which relied on information theory, as well as
the properties of charged rotating black holes in general
relativity\cite{bek}.

The next crucial insight came from Hawking while investigating quantum fields
in a black hole spacetime. He found that
external observers detect the emission of thermal radiation from
a black hole with a temperature proportional to its surface gravity,
$\kappa$ \cite{radi}:
\beq
k_{\scriptscriptstyle B}T={\hbar\kappa\over2\pi c}\ \ .
\label{temp}
\eeq
The surface gravity may be thought of as the redshifted acceleration
of a fiducial observer moving just outside the horizon\cite{wtext}.
In the simplest (\ie spherically symmetric) case, $\kappa=c^4/(4GM)$ where
$M$ is the mass of the black hole.

Previously, extensive studies of Einstein's equations
had culminated in the formulation of four laws of
black hole physics\cite{barcar}.
Hawking's discovery of the thermal radiance of black holes
was the key to realizing that these results were the laws
of thermodynamics applied to black holes.
In the present discussion, our primary interest will be
in the First and Second Laws.
The First Law of black hole mechanics takes the form
\begin{equation}
\frac{c^2\kappa}{8\pi G}\delta A=c^2\delta M-\Omega\,\delta J
\label{first}
\end{equation}
where $A$, $M$, $\Omega$ and $J$ are the horizon area, the mass,
the angular velocity (of the horizon), and angular momentum
of the black hole\cite{footgk}.
Here, $Mc^2$ is naturally identified with
the internal energy of the black hole, and $-\Omega\,\delta J$
appears to be a work term. Thus given the relation of the
surface gravity to the black hole temperature in Eq.~(\ref{temp}),
the identification of Eq.~(\ref{first}) with the usual thermodynamic
First Law (\ref{firsttherm}) is completed by recognizing that
the black hole entropy is \cite{radi}
\beq
S={k_{\scriptscriptstyle B}c^3\over\hbar\, G}\,{A\over 4}\ \  .
\label{entropy1}
\end{equation}
This formula applies for any black hole solution of Einstein's
equations\cite{foot1}. The Second Law of black hole thermodynamics
is then established by Hawking's area theorem, which states that
in any classical process involving black holes
and positive energy matter,
provided that naked singularities do not develop,
the total surface area of the event horizon will
never decrease\cite{areathm}. Thus one arrives at a consistent
framework for black hole thermodynamics by drawing upon results from
thermodynamics, quantum field theory and general relativity.
Much of the subsequent interest was and is motivated by the hope
that it may provide some insight into the nature of quantum gravity.

Now we would like to motivate studying the thermodynamic properties of
black holes within higher curvature gravity theories.
Whatever framework physicists eventually uncover to describe
quantum gravity, there should be a low energy effective
action which describes the dynamics of a ``background metric field''
for sufficiently weak curvatures at sufficiently long distances. On
general grounds, one expects that this effective gravity action will consist
of the classical Einstein action plus a series of
covariant, higher-dimension interactions
({\it i.e.,} higher curvature terms, and also higher derivative terms
involving all of the physical fields) induced by quantum effects.
While such effective
actions are typically pathological when considered as fundamental, they
may also be used to define mild perturbations for
Einstein gravity coupled to conventional
matter fields. It is within this latter context of Einstein
gravity ``corrected'' by higher dimension operators that we wish to
consider modifications of black hole thermodynamics.

Naive dimensional analysis suggests that the coefficients of all
higher dimension terms in such an effective lagrangian should
be dimensionless
numbers of order unity times the appropriate power of the Planck length.
Thus one might worry the effect of the higher dimension terms
would be the same order as those of quantum fluctuations, and so there
would seem to be little point in studying modifications to
{\it classical} black hole thermodynamics from higher dimension terms.
One motivation for
studying the classical problem is that it is of course
possible that the coefficients of some higher dimension terms are larger
than what would be expected from simple dimensional analysis. We would
like to know whether or not consistency with classical black hole
thermodynamics places any new restrictions on these coefficients.
Moreover, it is interesting to explore black hole thermodynamics in
generalized gravity theories in order to see whether the
thermodynamic ``analogy" is just a peculiar accident of Einstein gravity,
it is a robust feature of all generally covariant theories of gravity,
or it is something in between.

Certainly, Bekenstein's original considerations require only the
existence of an event horizon, but make no reference to the details
of the dynamics of the gravity theory which determines the precise
spacetime geometry. Hence these observations would be equally
applicable in higher-curvature gravity theories. Similarly,
the emission of Hawking radiation with a temperature as in Eq.~(\ref{temp})
is a result of quantum field theory for a spacetime containing a
horizon\cite{radi,field}. Again, this result is
independent of the details of any particular theory of
gravity.\footnote{Alternatively, this identification
follows from a semiclassical evaluation of the
partition function or the density of states in quantum gravity
with any action, as has been shown explicitly in the case of Einstein
gravity\cite{euclid,BY}.} Thus black holes in higher-curvature
theories will also emit thermal radiation with a temperature
proportional to the surface gravity.

{}From Euclidean path integral methods\cite{euclid}, it is clear that
a version of the First Law of black hole thermodynamics
still applies in any higher-curvature
theory. Applying these techniques to various specific theories
and specific black hole solutions, though, showed
that Eq.~(\ref{entropy1}) which equates the black hole entropy with
the surface area of the horizon no longer holds in
general\cite{failS}. Recently, the entropy was shown
to be given always by a local expression evaluated at the
horizon\cite{love,wald1,visser,wald2}.

Wald\cite{wald1} established the latter result very generally for any
diffeomorphism invariant theory via a new
(Minkowski signature) derivation
of the First Law of black hole mechanics.
For variations around a {\it
stationary} black hole background, he derived a modified First Law
\begin{equation}
\frac{\kappa}{2\pi c}\delta \S=c^2\delta M-\Omega\,\delta J
\label{firsta}
\end{equation}
where $\S$ is expressed as a {\it local} geometric density
integrated over a space-like cross-section of the horizon.
Since the black hole temperature is always given by Eq.~(\ref{temp})
independent of the details of the dynamics of the gravity theory,
the black hole entropy is naturally identified as
$S=(k_{\ssc B}/\hbar) \S$. If these expressions are truly to play the
role of an entropy, they should also satisfy the Second Law of
thermodynamics as a black hole evolves --- \ie
$\S$ should never decrease in any dynamical processes.
We have been able to establish this result at least within a
certain class of theories in which the curvature only enters
the action as powers of the Ricci scalar\cite{secondlaw}.

In the following, we begin with a brief review of Wald's derivation
of the First Law in section 2.
In section 3, we demonstrate that
the Second Law holds for arbitrary dynamical processes in a theory
where the gravitational Lagrangian is $R+\alpha R^2$.
In section 4, another proof of the Second Law is constructed
which follows closely Hawking's proof of the area theorem\cite{areathm}.
In fact, both of these proofs of the Second Law may be generalized
to include a larger class of theories where the gravitational action is a
polynomial of the Ricci scalar\cite{secondlaw}.
Section 5 presents a brief discussion of our results.
In the following, we adopt the standard
convention of setting $\hbar=c=k_{\scriptscriptstyle B}=1$.
Also, we employ the conventions of \cite{wtext} throughout, and we will
only consider asymptotically flat (four-dimensional) spacetimes.

\vglue 0.6cm
{\bf\noindent 2. Black Hole Entropy as Noether Charge}
\vglue 0.4cm
Wald\cite{wald1} constructed a new
derivation of the First Law
of black hole mechanics (\ref{firsta}) for any
theory which is invariant under diffeomorphisms
({\it i.e.,} coordinate transformations). In this construction,
the black hole entropy $S=\S$ is related to the Noether charge of
diffeomorphisms under the Killing vector field which generates the
event horizon in the stationary black hole background. Further,
$\S$ can always be expressed as a {\it local}\/ geometric density
integrated over a space-like cross-section of
the horizon\cite{wald1,onS}.
Thus the general result, Eq.~(\ref{firsta}), has in
common with the original First Law, Eq.~(\ref{first}),
the rather
remarkable feature that it relates variations in properties of the
black hole as measured at asymptotic infinity to a variation
of a geometric property of the horizon.
In the following, we will provide a brief introduction to Wald's techniques.
The interested reader is referred to Refs.~\cite{wald1,wald2,onS}
for complete details.

An essential element of Wald's approach is the
Noether current associated with diffeomorphisms\cite{wallee}.
Let $L$ be a Lagrangian built out of some set of
dynamical fields,
including the metric, collectively denoted as $\psi$.
Under a general field variation $\delta \psi$, the Lagrangian
varies as
\begin{equation}
\delta(\sqrt{-g} L)=\sqrt{-g}E\cdot\delta \psi + \sqrt{-g}\,\nabla_{\! a}
\theta^a(\delta \psi)\ \ ,
\label{dL}
\end{equation}
where ``$\cdot$" denotes a summation over the dynamical
fields including contractions of tensor indices, and $E=0$
are the equations of motion. With symmetry variations for which
$\delta(\sqrt{-g} L)
=0$, $\theta^a$ is the Noether current which is conserved
when the equations of motion are satisfied --- {\it i.e.,} $\nabla_{\! a}
\theta^a(\delta \psi)=0$ when $E=0$. Rather than vanishing
for diffeomorphisms, $\delta\psi=
{\cal L}_\xi\psi$,\footnote{${\cal L}_\xi\psi$ denotes the Lie
derivative of the field $\psi$ along the vector field $\xi^a$\cite{wtext}
--- \eg, ${\cal L}_\xi g_{ab}=\nabla_a\xi_b+\nabla_b\xi_a$.}
the variation of a covariant Lagrangian is a total derivative,
$\delta (\sqrt{-g} L)={\cal L}_\xi(\sqrt{-g} L)
=\sqrt{-g}\nabla_{\! a}(\xi^a\, L)$.
Thus one constructs a new Noether current,
\[
J^a=\theta^a({\cal L}_\xi \psi)- \xi^a L\ \ ,
\]
which satisfies $\nabla_{\!a}J^a=0$ when $E=0$.

A fact which is not well-appreciated is that for {\it any}\/ local symmetry,
there exists  a globally-defined Noether potential $Q^{ab}$,
satisfying $J^a=\nabla_{\! b}Q^{ab}$
where $Q^{ab}=-Q^{ba}$\ \cite{wald3}.
$Q^{ab}$ is a local function of the dynamical fields and a
linear function of the symmetry parameter ({\it i.e.,} $\xi^a$
in the present case). Of course,
this equation for $J^a$ is valid up to terms which vanish
when the equations of motion are satisfied.
Given this expression for $J^a$,
it follows that the Noether charge contained in a spatial volume $\Sigma$
can be expressed as a boundary integral $\oint_{\partial \Sigma}d^2\!x\,
\sqrt{h}\,\epsilon_{ab}Q^{ab}$, where $h_{ab}$ and
$\epsilon_{ab}$ are the induced metric and binormal form
on the boundary $\partial \Sigma$.

Wald's derivation of the First Law requires that the black hole
possess a bifurcate Killing horizon, which is defined as follows:
First, a Killing vector is a vector field generating an
invariance for a particular solution --- {\it i.e.,} ${\cal L}_\xi\psi=0$
for all fields. A Killing horizon is then
a null hypersurface whose null
generators are orbits of a Killing vector field.
If the horizon generators are geodesically
complete to the past (and if the surface gravity is nonvanishing),
then the Killing horizon
contains a space-like cross-section $B$, the {\it bifurcation surface},
on which the Killing field $\chi^a$
vanishes\cite{raczwald}.
Such a bifurcation surface is a fixed point of the Killing flow,
and lies at the
intersection of the two null hypersurfaces that comprise the
full Killing horizon.
For example, in the maximally extended Schwarzschild black hole
spacetime, the bifurcation surface is the two-sphere of area
$16\pi M^2$ at the origin of Kruskal $U$-$V$ coordinates.
The existence of bifurcate Killing horizons in general,
and its relation to the Zeroth Law (constancy of the
surface gravity), will be discussed in section 5.

The key to Wald's derivation of the First Law is the identity
\[
\delta H=\delta\int_\Sigma dV_a J^a -
\int_\Sigma dV_a\nabla_b(\xi^{a}\theta^{b}-\xi^{b}\theta^{a} ),
\]
where $H$ is the Hamiltonian generating evolution along
the vector field $\xi^a$, and
$\Sigma$ is a spatial hypersurface with volume element $dV_a$.
This identity is satisfied for arbitrary variations of the fields
away from any background solution.
If the variation is to another solution, then one can replace
$J^a$ by $\nabla_b Q^{ab}$, so the variation of the Hamiltonian
is given by surface integrals over the boundary $\partial\Sigma$. If,
moreover, $\xi^a$ is a Killing vector of the background solution, then
$\delta H=0$. In this case one obtains an identity relating the
various surface integrals over $\partial\Sigma$.

Suppose now that the background solution is chosen to be a
stationary black hole with horizon-generating Killing field
$\chi^a\partial_a={\partial\ \over\partial t}+\Omega{\partial\ \over
\partial\phi}$,
and the hypersurface $\Sigma$ is chosen to
extend from asymptotic infinity down
to the bifurcation surface where $\chi^a$ vanishes.
The surface integrals at infinity then yield precisely the
mass and angular momentum variations, $\delta M-\Omega\delta J$,
appearing in Eq.~(\ref{firsta}),
while the surface integral at the bifurcation surface reduces to
$\delta\oint_Bd^2\!x\,\sqrt{h}\,\epsilon_{ab}Q^{ab}({\chi})$.
Finally, it can be shown that the latter surface integral always
has the form $(\kappa/2\pi) \delta \S$, where
$\kappa$ is the surface gravity of the background black hole, and
$\S=2\pi\oint_Bd^2\!x\,\sqrt{h}\,\epsilon_{ab} Q^{ab}({\tilde{\chi}})$,
with $\tilde{\chi}^a$, the Killing vector scaled to have unit
surface gravity.

By construction $Q^{ab}$ involves the Killing
field $\tilde{\chi}^a$ and its derivatives. However, this
dependence can be eliminated as follows\cite{wald1}. Using
Killing vector identities, $Q^{ab}$ becomes a function of only $\tilde{\chi}^a$
and the first derivative, $\nabla_{\! a}\tilde{\chi}_b$.
At the bifurcation surface, though, $\tilde{\chi}^a$ vanishes and
$\nabla_{\! a}\tilde{\chi}_b={\epsilon}_{ab}$, where ${\epsilon}_{ab}$
is the binormal to the bifurcation surface.
Thus, eliminating the term linear in $\tilde{\chi}^a$ and replacing
$\nabla_{\! a}\tilde{\chi}_b$ by ${\epsilon}_{ab}$ yields a completely
geometric
functional of the metric and the matter fields, which may be denoted
$\tilde{Q}^{ab}$. One can show that the
resulting expression,
\begin{equation}
\S=2\pi\oint d^2\!x\,\sqrt{h}\,\epsilon_{ab}\, \tilde{Q}^{ab},
\label{Sgeom}
\end{equation}
yields the
correct value for $\S$ when evaluated not only at the
bifurcation surface, but in fact on
an arbitrary cross-section of the Killing horizon\cite{onS}.
Thus this latter expression is a
natural candidate for the entropy of a general non-stationary
black hole.

Using Wald's technique,
the formula for black hole entropy has been found
for a general Lagrangian of the following form:
\[
L=L(\psi,\nabla_{\! a}\psi,g_{ab}, R_{abcd}, \nabla_e R_{abcd},
\nabla_{(e_1}\nabla_{e_2)} R_{abcd}, \ldots)\ \ ,
\]
that is involving only first derivatives of the matter fields
(denoted by $\psi$), and the Riemann tensor and symmetric derivatives
of $R_{abcd}$ up to some finite order $n$. $\S$ may then be
written\cite{visser,onS,wald2}
\beq
\S=-2\pi\oint d^2\!x\sqrt{h}\ \sum^n_{m=0} (-)^m\
\nabla_{(e_1}\ldots\nabla_{e_m)}Z^{e_1\cdots e_m:abcd}\
\epsilon_{ab}\,{\epsilon}_{cd}
\label{wform}
\eeq
where the $Z$-tensors are defined by
\[
Z^{e_1\cdots e_m:abcd}\equiv {\partial\, {L}\over
\partial\nabla_{(e_1}\ldots\nabla_{e_m)} R_{abcd}}\ \ .
\]
As a more explicit example, consider the action
\[
I={1\over16\pi G}\int d^4\!x\,\sqrt{-g}\left(R+\alpha R^2+\beta R_{ab}R^{ab}
\right)
\]
for which one finds
\beq
\S={1\over4G}\oint d^2\!x\,\sqrt{h}\left(1+2\alpha R+\beta(R-h^{ab}R_{ab})
\right)\ \ .
\label{results}
\eeq
Here, the first term yields the expected contribution for Einstein
gravity, namely $A/(4G)$. Thus just as the Einstein term in the action
is corrected by higher-curvature terms, the Einstein
contribution to the black hole entropy receives higher-curvature corrections.

Note that the present description of Wald's construction has neglected
certain important
details. For example, the reader is directed to Ref. \cite{wald2}
for an explanation of
why the elimination of explicit dependence of $\S$ on the
Killing field (discussed in the paragraph containing Eq.~(\ref{Sgeom}))
is valid even for when variations to non-stationary solutions are allowed.
Further, a number of ambiguities arise in the construction
of $\tilde{Q}^{ab}$\cite{onS}, so Eqs. (\ref{wform}) and (\ref{results})
should be understood as the result of making certain (natural) choices
in the calculation.
None of these ambiguities have any effect when the charge
is evaluated on a stationary horizon\cite{onS}, but they
will become significant for non-stationary horizons.
In this case, a choice which yields an entropy that satisfies
the second law would be a preferred definition\cite{secondlaw}.
\vglue 0.6cm
{\bf \noindent 3. The Second Law in one example}
\vglue 0.4cm

Given that the identification of the surface gravity with the Hawking
temperature in Eq.~(\ref{temp}) is universal,
Wald's generalized First Law has the natural interpretation as the
First Law of thermodynamics for higher-curvature theories
with the black hole entropy $S=\S$. If the latter
expressions are truly to play the role of an entropy, they should
also satisfy the second law of thermodynamics as a black hole
evolves --- {\it i.e.,}
$\S$ should never decrease in any dynamical processes.

For general relativity with $S=\S= A/(4G)$, the Second Law is
established by Hawking's area theorem\cite{areathm}.
An essential ingredient in the proof of this theorem is the assumption
that the null convergence condition $R_{ab}k^ak^b\ge0$ holds for all null
vectors $k^a$. This condition is implied by
the Einstein field equation
\beq
R_{ab}-{1\over2}g_{ab}R=8\pi G\, T_{ab}
\label{eom}
\eeq
together with the  null energy condition for the matter
stress-energy, \ie
\beq
T_{ab}\,k^a k^b\ge0\qquad{\rm\ for\ any\ null\ vector\ }k^a\ \ .
\label{null}
\eeq
Another essential ingredient is cosmic censorship --- \ie it is
assumed that
naked singularities do not develop in the processes of interest.

In theories where higher curvature interactions are
included along with the Einstein Lagrangian, the equations of motion
may still be written in the form of Eq.~(\ref{eom}) if the contributions
from the higher curvature interactions are included in the stress-energy
tensor. However, these contributions typically spoil
the energy condition required to prove the area theorem. Hence, in these
theories, one can not establish an area increase theorem,
but this result is not relevant
since the entropy is no longer identified with the area
in such a case.
The relevant question is whether
or not the quantity $\S$ appearing in the First Law (\ref{firsta})
satisfies an increase theorem.
If so, one would have a
Second Law of black hole thermodynamics for these theories,
further
validating the interpretation of $\S$ as the black hole entropy.

In this section we establish the Second Law for arbitrary
dynamical processes involving black holes in the theory
given by a higher curvature action of the form
\beq
I_{\ssc 0}=\int \ddx\sqrt{-g}\left[{1\over16\pi G}(R+\al R^2)+L_m(\psi,g)
\right]
\label{act0}
\eeq
where $L_m$ denotes a conventional Lagrangian for some collection of matter
fields, denoted $\psi$. The latter Lagrangian will also
contain the metric, but we assume that it contains no derivatives of
the metric. As in Eqs.~(\ref{wform}) or (\ref{results}), Wald's black hole
entropy can be written\cite{visser,wald2,onS}
\beq
\S={1\over4G}\int_\ch \dtx\, \sqrt{h}\, (1+2\al\, R)
\label{entrope}
\eeq
where the integral is taken over a space-like cross-section of the horizon,
$\ch$.

The gravitational field equations arising from the action (\ref{act0})
are
\beqa
R_{ab}-{1\over2}g_{ab}R&=&8\pi G\, T^m_{ab}(\psi,g)+2\alpha\,
(\,\nabla_a\nabla_bR-g_{ab}\nabla^2R
\nonumber\\
&&\qquad\qquad-\ R\,R_{ab}+{1\over4}g_{ab}\,R^2\,)\ \ .
\label{movie}
\eeqa
We will assume that
matter stress-energy tensor, $T^m_{ab}=-{2\over\sqrt{-g}}{\delta\sqrt{-g}
L_m\over\delta g^{ab}}$, does satisfy the null energy condition (\ref{null}).
However, if one treats the entire expression on the right hand side of this
equation as the stress-energy tensor of Einstein's equations (\ref{eom}),
it is clear that this total stress-energy does not satisfy the
null energy condition (\ref{null}) because of the higher curvature
contributions (\ie the terms proportional to $\al$). Thus, as
discussed above, Hawking's proof of the area
theorem does not apply here. What is desired, though, is to establish an
increase theorem for $\S$ in Eq.~(\ref{entrope}).
Our approach will be the following. First, we show that the present higher
curvature theory is equivalent to Einstein gravity
for a conformally related metric coupled to
an auxiliary scalar field, as well as to the original matter fields.
Second, we argue that
the black hole entropy in the higher curvature theory is identical to that
in the conformally related theory.
Finally, since Hawking's area theorem holds in the Einstein-plus-scalar
theory, we conclude
that the entropy never decreases in the original
theory (\ref{act0}).

The equivalence of the higher curvature theory (\ref{act0}) to Einstein
gravity coupled to an auxiliary scalar field has been discussed previously
by many authors\cite{conform}. The first step
is to introduce a new scalar field $\phi$, and a new action,
which is linear in $R$,
\beq
I_{\ssc 1}=\int \ddx\sqrt{-g}\left\{{1\over16\pi G}\left[(1+2\al\phi)R
-\al\phi^2\right]+L_m(\psi,g)\right\}\ \ .
\label{act1}
\eeq
The $\phi$ equation of motion is simply $\phi= R$, and
one recovers the original action upon substituting this equation into
Eq.~(\ref{act1}) --- \ie $I_{\ssc 1}(\phi= R)=I_{\ssc 0}$.
In the form of Eq.~(\ref{act1}), the action contains no terms that
are more than quadratic in derivatives. However, this action does
contain an unconventional interaction, $\phi R$.
Hence in the metric equations of motion,
\beqar
R_{ab}-{1\over2}g_{ab}\,R&=&8\pi G\, T^m_{ab}(\psi,g)+2\alpha\,
(\,\nabla_a\nabla_b\phi-g_{ab}\nabla^2\phi\\
&&\qquad\qquad-\ \phi\,R_{ab}+{1\over2}g_{ab}\,\phi R
-{1\over4}g_{ab}\,\phi^2\,)\ ,
\eeqar
the total stress-energy tensor appearing on the right hand side
still contains some problematic contributions (\eg $\nabla_a\nabla_b\phi$),
which prevent the null energy condition from being satisfied.

The $\phi R$ interaction can be removed by performing the
following conformal transformation
\beq
{g}_{ab} = (1+2\al\phi)^{-1}\bar{g}_{ab}\ \ .
\label{two}
\eeq
In terms of $\bar{g}_{ab}$, the action (\ref{act1}) becomes
\beqa
I_{\ssc 2}&=&\int \ddx\sqrt{-\bar{g}}\left\{{1\over16\pi G}\left[\bar{R}
-{3\over2}\left({2\al\over1+2\al\phi}\right)^2\bar{\nabla}_a\phi
\bar{\nabla}^a\phi-{\al\phi^2\over(1+2\al\phi)^{2}}\right]\right.
\nonumber\\
&&\qquad\qquad\qquad\qquad\qquad
\left.\vphantom{\left[\left({2\al\over1+2\al\phi}\right)^2\right]}
+(1+2\al\phi)^{-2}L_m(\psi,(1+2\al\phi)^{-1}\bar{g})
\right\}\ ,
\label{act2}
\eeqa
which includes the standard Einstein-Hilbert action for $\bar{g}_{ab}$
and the auxiliary scalar $\phi$ with less conventional
couplings --- see below. The $\bar{g}_{ab}$ equations of motion are now
\beqa
\bar{R}_{ab}-{1\over2}\bar{g}_{ab}\,\bar{R}&=&{8\pi G\over1+2\al\phi}
T^m_{ab}(\psi,(1+2\al\phi)^{-1}\bar{g})+
{3\over 2}\left({2\al\over1+2\al\phi}\right)^2
\bar{\nabla}_a\phi\bar{\nabla}_b\phi
\nonumber\\
&&\qquad\qquad-\ {1\over2}\bar{g}_{ab}
\left[{3\over 2}\left({2\al\over1+2\al\phi}
\right)^2\bar{\nabla}_{\!c}\phi
\bar{\nabla}^c\phi + {\al\phi^2\over(1+2\al\phi)^{2}}\right]\ \ .
\label{moving}
\eeqa
The most important feature of this final theory for our purposes
is that, assuming $1+2\alpha\phi>0$,
the total stress-energy tensor appearing on the right hand
side above satisfies the null energy condition (\ref{null}).

Given the absence of higher derivative or unconventional gravity
couplings, the black hole entropy for $I_2$
is given by $\bar{\S}=\bar{A}/(4G)$,
just as for Einstein gravity. Since the equations of motion (\ref{moving})
are Einstein's equations, and the null energy condition
is satisfied by the stress-energy tensor,
Hawking's proof of the area theorem is valid for the $I_2$ theory
with the assumptions that cosmic censorship holds
for $\bar{g}_{ab}$  and that $1+2\alpha\phi>0$.
(The latter assumption will be further discussed below.)
Hence there is a classical entropy increase theorem for
the theory defined by $I_2$ in Eq. (\ref{act2}).

Now Eq.~(\ref{two}) along with $\phi=R$ provides a
mapping between the solutions for the
Einstein-plus-scalar theory defined by $I_{\ssc 2}$,
and the original higher curvature theory defined by $I_{\ssc 0}$,
in which the metrics are related by a conformal transformation
\beq
\bar{g}_{ab} = (1+2\al R)g_{ab}\ \ .
\label{map}
\eeq
The conformal transformation (\ref{map})
preserves the causal structure of the solutions and,
if $g_{ab}$ is asymptotically flat, then so is
$\bar{g}_{ab}$. Thus, if $g_{ab}$ is an asymptotically flat black hole,
then so is $\bar{g}_{ab}$, and they have the same horizon and
surface gravities \cite{tjgk}. Furthermore, since the asymptotic
forms of $g_{ab}$ and $\bar{g}_{ab}$ agree, the mass and angular
momenta of the two spacetimes agree.
Further, the angular velocities agree
on stationary horizons, because the horizon generating null combination of the
time translation and rotation Killing fields is preserved
under the conformal transformation.

In short, we have shown all of the ingredients, other than the entropy, in the
First Law (\ref{firsta}) agree. Thus, for all variations,
the changes in the entropies must also agree. Therefore the
entropies themselves are equal to within a constant in each connected class
of black hole solutions.
Since the area increase
theorem for the Einstein-plus-scalar theory gives
$\delta\bar{\S}\ge0$ in any dynamical process,
we conclude that $\delta{\S}\ge0$ for the corresponding
process in the higher curvature theory.
We have thus established a classical Second Law in the higher curvature
theory defined by the action $I_0$ in Eq. (\ref{act0}).

A number of remarks on the above analysis will now be made.
Using the conformal relation (\ref{map}) between the
two metrics, the ``barred" entropy can be expressed directly
in terms of
$g_{\mu\nu}$:
\beq
\bar{\S}(\bar{g})={1\over4G}\int_{\bar{\ch}}\dtx\,\sqrt{\bar{h}}
={1\over4G}\int_{\bar{\ch}}\dtx\,\sqrt{h}\,(1+2\al\,R)\ \ .
\label{mapent}
\eeq
For stationary black holes this agrees with the
entropy in the higher curvature theory (\ref{entrope}) as determined
directly from the First Law in that theory (it was already argued above that
$\bar{\ch}$ corresponds to a cross-section
of the event horizon for the metric $g_{ab}$ as well). This agreement
is explained by the reasoning given in the preceding paragraph.
It should be emphasized however that the First Law, which applies
to variations away from a stationary black hole background,
does not determine the form of the entropy during arbitrary
dynamical processes \cite{wald1,onS}.

In presenting the result (\ref{entrope})
for the black hole entropy as determined by the First Law, we
chose the simplest geometric formula which naturally extends to
a dynamical horizon.
Here we have shown that the Second Law is obeyed,
even {\it during} a dynamical process,  by the entropy given
by that particular formula, reproduced in Eq.~(\ref{mapent}).
The relation (\ref{map}) gives an unambiguous result for the
dynamical entropy and so
can be used to resolve
the ambiguities \cite{wald2,onS} inherent in the
Noether charge construction of \cite{wald1}.
In the present case, the alternate proposal
for dynamical entropy of Ref.~\cite{wald2},
which uses a boost invariant projection,
gives a result for the dynamical entropy that differs from Eq.~(\ref{mapent})
for non-stationary black holes.
Unless there are two different entropy functionals obeying
a local increase
law, it appears that the proposal of Ref.~\cite{wald2} is inconsistent
with the Second Law
during the dynamical process in the present theories.

There is one considerable assumption in preceding discussion
which we have not yet addressed. For the mapping between the
solutions of the two theories (\ref{map}) to exist,
it is necessary that
the factor $1+2\al R$ is positive.
(Note that in asymptotically flat regions
this factor tends to one.)
Thus, given a black hole solution of the higher curvature
theory, one must have $R>-{1\over2\al}$ (for positive $\al$)
everywhere outside of the black hole and on the event horizon.

{}From the point of view of the Einstein-plus-scalar theory,
one requires that $\phi>-{1\over2\al}$ everywhere outside of the event
horizon for the mapping to a solution of the higher curvature
theory to exist. Recall that cosmic censorship was assumed in the proof
of the area increase theorem for this theory. This assumption
rules out dynamical processes in which a black hole begins with
a configuration with $\phi>-{1\over2\al}$ everywhere initially, and
evolves to one with $\phi\le-{1\over2\al}$ in some region outside
of the horizon. The reason is that, by the equations of motion
(\ref{moving}) when $\phi=-{1\over2\al}$, the stress-energy tensor
is singular and hence the Einstein tensor,
$\bar{R}_{ab}-{1\over2}\bar{g}_{ab}\bar{R}$, is singular\cite{footsing}.
Cosmic censorship
for $\bar{g}_{ab}$
would only allow such curvature
singularities to develop
behind the event horizon, and hence rules out any process in which
$\phi=-{1\over2\al}$ is reached outside of the horizon.

An alternative argument showing that it is consistent to make the
restriction $\phi>-{1\over2\alpha}$ can be given by considering the character
of the potential term in the Einstein-plus-scalar theory\cite{andy}.
The non-standard kinetic term for the scalar
field $\phi $ in Eq. (\ref{act2}) can be replaced with an ordinary one
by defining a new scalar field $\varphi :=
\beta^{-1}\ln (1+2\alpha \phi )$, where $\beta =\sqrt{16\pi
G/3}$. In terms of $\varphi$ the action $I_2$ takes the form
\begin{equation}
I_{\ssc 3}=\int \ddx \sqrt{-\bar{g}}[ \frac{1}{16\pi G}
\bar{R}-\frac{1}{2}\bar{\nabla }_a\varphi \bar{\nabla }^a
\varphi - V(\varphi ) + e^{-2\beta \varphi }
L_m(\psi ,e^{-\beta \varphi }\bar{g})],
\label{act3}
\end{equation}
where $V(\varphi )=\frac{1}{64\pi G\alpha }e^{-2
\beta \varphi }(e^{\beta \varphi }-1)^2 $.
Now the singular point $\phi =-\frac{1}{2\alpha }$ corresponds to $\varphi
\rightarrow -\infty $. Provided $\alpha>0$, the potential $V(\varphi )$
rises exponentially as $\varphi\rightarrow -\infty$.
The term involving the matter Lagrangian has the same exponential for
a prefactor, and so one may worry that it may undermine the barrier
due to $V(\varphi)$. The kinetic terms for the matter fields
will include at least one inverse metric which will
bring the rate of exponential growth down by a factor of
$\exp(\beta\varphi)$ for these contributions.
 We will assume that any matter
potential is non-negative --- this amounts to extending the
null energy condition for $T^m_{ab}$ to the dominant energy
condition --- so that these terms can only increase the potential
barrier as $\varphi\rightarrow-\infty$.
Thus, as long as the metric and matter fields
do not become singular, the dynamics of $\varphi$
as $\varphi\rightarrow -\infty$ will be dominated by
the potential barrier so $\varphi$ will not run off to $-\infty$.
Therefore, initial data satisfying the bound
$\phi>-{1\over2\alpha}$ will evolve within the bound, as long as the other
fields remain nonsingular.

Note that the argument just given breaks down if $\alpha<0$ since the
potential is then exponentially {\it falling} as $\varphi\rightarrow -\infty$.
In fact, the theory is unstable for negative $\alpha$. The previous argument
did not seem to require that $\alpha$ be positive, however it did assume cosmic
censorship which, presumably, would be violated in the unstable
theory with $\alpha<0$.

Given the above arguments that
$\phi=-{1\over2\al}$ is never reached outside
the event horizon for positive $\alpha$,
one may also rule out processes in the higher curvature theory
in which a black hole evolves to reach $R=-{1\over2\al}$ somewhere
outside of the event horizon. From a superficial
examination of the higher curvature equations of motion (\ref{movie}),
$R=-{1\over2\al}$ does not appear to be singular.
However, there
is no obstruction to mapping the initial part of the evolution to
the Einstein-plus-scalar theory, where it becomes a process
leading up to a naked singularity
of $\bar{g}_{ab}$ at the point where $\phi=-{1\over2\al}$,
as discussed above. Such processes were ruled out by the assumption
of cosmic censorship for $\bar{g}_{ab}$, though,
hence we have also ruled out the
corresponding evolution in the higher curvature theory.

\vglue 0.6cm
{\bf \noindent 4. Direct proof of Second Law}
\vglue 0.4cm

Our method of establishing the Second Law above
used the fact that the higher curvature gravity theory is conformally
related to Einstein gravity in which the area theorem holds.
This special feature of these theories is not shared by most general
higher curvature theories, so it would be most interesting to see how the
Second Law could be established {\it directly},
without making
use of the conformal transformation technique. In the present section we
will construct such a direct proof for the curvature
squared theory studied in the previous section. The exercise will be
instructive for efforts to establish entropy increase theorems for theories
that are not susceptible to the conformal transformation ``trick".

 Suppose that the black hole entropy of a gravity theory takes
the following form
\begin{equation}
S = \frac{1}{4G}\int_{\cal H}\dtx\, \sqrt{h}\, e^\rho,
\label{expent}
\end{equation}
where $e^\rho$ is a scalar function of the local geometry at the horizon.
For the theory considered in the preceding section  one has
$e^\rho=1+2\alpha R$. The method to be used here will rely critically on
the fact that $e^\rho$ is necessarily positive, and $\rho=0$
when the curvature vanishes.

We wish to consider the change of this entropy along the null
congruence generating the event horizon under any dynamical evolution.
Let $k^a$ be the null tangent vector field of the horizon generators with
respect to the affine parameter $\lambda $.
Then one has
\begin{equation}
S'=\frac{1}{4G}\int_{\cal H}\dtx \sqrt{h}\, e^\rho\, \tilde{\theta}
\end{equation}
with
\begin{equation}
\tilde{\theta}:= \theta +\partial_\lambda\rho,
\end{equation}
where
$\theta=d(\ln\sqrt{h})/d\lambda =\nabla_ak^a$
is the expansion of the horizon generators.

\def\pl{\partial_\lambda}
Now the question
is whether or not there can exist a point along the null geodesics at
which $\tilde{\theta }$ becomes negative. In order to answer this
question, we use the Raychaudhuri equation, as in the proof of area theorem,
to obtain an expression for $\pl\tilde{\theta}$:
\begin{eqnarray}
\pl\tilde{\theta}&=& \pl\theta+\pl^2\rho
\nonumber     \\
&=& -\textstyle{\frac{1}{2}}{\theta }^{2}-
{\sigma }^2 - k^ak^bR_{ab} + k^ak^b\nabla_a\nabla_b\rho,
\end{eqnarray}
where $\sigma^2$ is the  square of the shear.

For the $R+\al R^2$ theories, it is easy to see that the equations of
motion (\ref{movie}) imply that
$k^ak^b(R_{ab}-\nabla_a\nabla_b\rho)=(8\pi G\,e^{-\rho} T^m_{ab}
+\nabla_a\rho\nabla_b\rho)\,k^ak^b$,
which is non-negative provided the null energy condition holds for the
matter fields. Thus in those theories one has
\begin{equation}
\pl\tilde{\theta }
\leq -\textstyle{\frac{1}{2}} {\theta }^2,
\end{equation}
or
\begin{equation}
\pl[\tilde{\theta}^{-1}]
\ge\frac{1}{2} (\theta/\tilde{\theta})^2.
\label{ineq}
\end{equation}

Now we follow Hawking's proof of the area theorem, with
$\tilde{\theta}$ in place of $\theta$.
Suppose at some point on the horizon we have $\tilde{\theta}<0$.
Then in a neighborhood of that point one can deform a space-like slice of
the horizon slightly outward to obtain a compact space-like
two-surface $\Sigma$ so that $\tilde{\theta}<0$ everywhere on $\Sigma$,
$\tilde{\theta}$ being
defined along the outgoing null geodesic congruence orthogonal to $\Sigma$.
If cosmic censorship is assumed, then there is necessarily some
null geodesic orthogonal to $\Sigma$ that remains on the boundary
of the future of $\Sigma$ all the way out to ${\cal I}^+$\cite{areathm}.
Asymptotic flatness (where components of the Riemann tensor in an
orthonormal frame all fall off at least as $r^{-3}$)
implies that $\rho\rightarrow 0$ like $\lambda^{-3}$
at infinity, whereas $\theta$ goes like $\lambda^{-1}$, where $\lambda$
is the affine
parameter along an outgoing null geodesic. Therefore
$\theta/\tilde{\theta}\rightarrow 1+O(\lambda^{-3})$,
so the inequality (\ref{ineq}) implies that, as one follows the geodesic
outwards from $\Sigma$, $\tilde{\theta}$ reaches $-\infty$
at some finite affine parameter.
Since $\tilde{\theta}=\theta+\pl\rho$, this means that either
$\theta$  or $\pl\rho$ goes to  $-\infty$. In the former case we have a
contradiction, as in the area theorem, since it implies there is a
 conjugate point on the geodesic, which cannot happen since the
geodesic stays on the boundary of the future of $\Sigma$ all the way out to
${\cal I}^+$. In the latter case we have a naked singularity,
since $\pl\rho=e^{-\rho}\pl e^\rho$, and $e^\rho$ was assumed from
the beginning in (\ref{expent}) to be a nonvanishing function
of the curvature.

For the present theory, we have in particular
$\pl\rho=k^a\nabla_a \rho=2\al(1+2\al R)^{-1}k^a\nabla_aR$.
In the preceding section we argued
that $1+2\al R$ never goes to zero outside
the horizon in the case of a stable theory,
so the divergence of $\pl\rho$ implies a divergence of
$k^a\nabla_a R$,
but this divergence also violates cosmic censorship.
Therefore we conclude that cosmic censorship and the null
energy condition
on the matter imply that the black hole entropy (\ref{expent})
can never decrease for the stable theories.
Note that in making this argument we have used the condition
$1+2\al R>0$ that was established via the conformal transformation
trick, so we do not quite have a fully ``direct proof" of the
Second Law.

The above argument suggests that the ``weakest" naked singularity
which might create
a violation of the Second Law would be a divergence in $k^a\nabla_a R$.
It seems that this could happen even if the curvature itself is nonsingular
everywhere. However, if one imposes also the equations of motion of the
theory, then a divergence in $k^a\nabla_a R$ would necessarily entail also
a divergence in either the curvature tensor or the matter stress tensor.

\vglue 0.6cm
{\bf \noindent 5. Discussion}
\vglue 0.4cm

Present investigations indicate quite strongly that black
hole thermodynamics is a robust feature of all generally covariant
theories of gravity. In particular, Wald's construction provides
a general First Law of black hole mechanics
\[
\frac{\kappa}{2\pi}\delta \S=\delta M-\Omega\,\delta J
\]
for any such theory.
The principle difference between separate theories
is the precise formula for  the black hole entropy, $\S$.
In general, though, $\S$ is expressed as a {\it local} geometric
density integrated over a cross-section of the horizon. One might
have expected the latter result following Bekenstein's original observations.

To even state the First Law presumes the validity of the Zeroth Law,
namely, that the surface gravity or temperature is constant across a
stationary event horizon. In Einstein
gravity the Zeroth Law
for a Killing horizon can be proved if the
dominant energy condition is assumed\cite{barcar}.
It can be similarly
established for the $R+\alpha R^2$ theory using the conformal
transformation technique described in section 3. One need only
notice that the stress energy tensor for the action $I_2$ (\ref{act2})
or $I_3$ (\ref{act3}) satisfies the dominant energy condition,
assuming that $\alpha$ is positive, and that the original matter
fields satisfy the dominant energy condition.
Hence the surface gravity
of any Killing horizon for the $\bar{g}_{ab}$ metric must be constant
and,
since surface gravity is conformally invariant\cite{tjgk},
the same must be true of the Killing horizons for ${g}_{ab}$.
In more general higher curvature theories the validity of the
Zeroth Law remains an important open question.

It is worth emphasizing that unless the event horizon
is a Killing horizon,
the abovementioned proofs of the Zeroth Law are not applicable.
In Einstein gravity, Hawking proved that the
event horizon of a stationary black hole must be a Killing
horizon\cite{areathm}.
To our knowledge, this proof has not been extended to general higher
curvature theories, or even to higher dimensional Einstein gravity.
For the higher curvature theories
considered in this paper,
at least in four dimensions, it seems likely that
Hawking's proof can be imported via the conformal transformation relating
the theory to Einstein gravity with matter.
For stationary black
holes in more general higher curvature theories,
whether or not stationary event horizons are necessarily Killing
horizons is another important open question.

Recall that Wald's derivation of the First Law required a Killing horizon
with a bifurcation surface.
Although not all Killing horizons have bifurcation surfaces,
those that satisfy the Zeroth Law do\cite{raczwald}. More precisely,
if the surface gravity is constant and nonvanishing
on a slice of a Killing
horizon, then there exists a stationary extension
of the spacetime
(at least in a neighborhood of the
horizon) to one that includes a bifurcation surface.
Moreover given a slice of a Killing horizon where
the surface gravity is {\it not} constant,
then a similar extension exists but
there is necessarily a curvature singularity where the
bifurcation surface would have been. Conversely, if
a Killing horizon has a regular bifurcation surface, then
the surface gravity is necessarily constant.
The Zeroth Law thus holds for {\it any} bifurcate Killing horizon,
irrespective of the underlying gravitational dynamics.
However, since one has no independent proof that  Killing horizons
in a given theory necessarily possess a regular bifurcation surface,
this does not help one to establish the Zeroth Law
in general.

In sections 3 and 4, we presented two proofs of the Second Law for
the higher curvature theory in Eq.~(\ref{act0}). For this
theory,
\[
\S={1\over4G}\oint\dtx\sqrt{h}(1+2\al R)
\]
will never decrease in any classical process involving black holes.
In fact, both of these proofs (as well as the proof of the Zeroth Law)
are easily generalized for higher
curvature theories\cite{secondlaw} with actions of the form
\[
I_{\ssc 0}=\int \ddx\sqrt{-g}\left[{1\over16\pi G}
(R+P(R))+L_m(\psi,g)\right]
\]
where $P$ is a polynomial in the Ricci scalar,
$P(R)=\sum_{n=2}a_n R^n$.
For these theories, the black hole entropy is
\beq
\S={1\over4G}\oint\dtx\sqrt{h}\,(1+P'(R))
\label{entbb}
\eeq
where $P'(R)=\sum_{n=2}na_n R^{n-1}$.
Similarly to the $R^2$ theory, these proofs of the Second Law require
that the matter fields satisfy the null energy condition,
and that the coupling constants $a_n$
appearing in $P(R)$ be restricted such that the slope of the entropy
density $1+P'(R)$ (plotted versus $R$) is positive
for positive $R$, and also between $R=0$ and the first negative
value of $R$ where $1+P'(R)$ vanishes.\cite{secondlaw}

As for the $R^2$ example discussed in detail, \ie $P(R)=\al R^2$,
the expression $1+P'(R)$ must be positive in order
to implement the
conformal transformation between the original higher curvature theory
and the Einstein-plus-scalar theory, and also to ensure the
null energy condition is satisfied in the latter theory.
This positivity
is also an essential ingredient for a direct proof, as in section 4.
It is interesting that
precisely the same expression plays the role of the entropy
surface density in Eq.~(\ref{entbb}). Thus the positivity restriction
translates on the horizon to the condition that the local entropy
density should be positive everywhere.
In particular it leads to
the total black hole entropy always being positive.
The latter is a minimum requirement that must be fulfilled
if this entropy is to have a statistical mechanical origin.
The fact that we actually require
a {\it local} positivity condition on the entropy density
is suggestively consistent with the idea that this density may have a
statistical interpretation.
In any event, these (and other higher curvature) theories
may provide a more refined test of the various proposals to
explain the statistical origin of black hole entropy.

One would like to extend the Second Law of black hole
thermodynamics to more
general higher curvature theories. One extension which we have found
is to consider general higher curvature theories but restricting
the black hole evolution to quasi-stationary processes\cite{secondlaw}.
For such processes in which a (vacuum) black
hole accretes positive energy
(neutral) matter --- \ie $T^m_{ab}\ell^a\ell^b\ge0$
for any null vector $\ell^a$ --- the Second Law is a direct consequence
of the First Law of black hole mechanics, independent
of the details of the gravitational action.

The investigations presented in sections 3 and 4\cite{secondlaw}
have been limited
to an intrinsic or classical Second
Law --- \ie we have only dealt with the increase of the black
hole entropy alone. In general relativity, we know
that the effective transfer of negative energy from quantum fields to
a black hole can lead to a decrease in the horizon entropy
(\ie horizon area), and the same is true
for these higher curvature effective theories since
black holes still produce Hawking radiation in these theories.
Thus it is important to ask whether a generalized
Second Law ($\delta (S_{BH}+S_{outside})\ge 0$) also holds.
In general relativity, there are arguments that the
generalized Second Law applies for quasi-stationary
processes involving positive energy matter\cite{gsl}.
These arguments seem to carry over to stable
higher curvature gravity theories as well,
since they do not involve the equations of motion but rather lean on the
First Law and the maximum entropy property of thermal radiation.
The validity of the generalized Second Law
for dynamical processes that are {\it not} quasi-stationary remains
an important open question.

\vglue 0.6cm
{\bf \noindent Acknowledgements \hfil}
\vglue 0.4cm
We would like to acknowledge useful discussions with
J. Friedman, V. Iyer, R.M. Wald and R. Woodard during the
course of our work.
R.C.M.\ was supported by NSERC of Canada, and Fonds FCAR du
Qu\'{e}bec.  T.J.\ and G.K.\ were supported in part by NSF Grant~PHY91--12240.
We would also like to thank the ITP, UCSB for their
hospitality during the early stages of this work, where the
research was supported by NSF Grant~PHY89--04035.
T.J.\ would like to thank the Institute for Theoretical Physics
at the University of Bern for hospitality and Tomalla Foundation
Zurich for support during the summer of 1994.

\vglue 0.6cm
{\bf\noindent References \hfil}
\vglue 0.4cm


\begin{thebibliography}{99}
\bibitem
{footgk} In the present discussion, we will only consider uncharged
black holes in four dimensions. The extension to include electrically
charged or higher dimensional cases is straightforward.
\bibitem
{footsing}
We assume that the singularities in the individual terms in the
stress-energy tensor of Eq.~(\ref{moving}) do not cancel to
avoid a curvature singularity.
\end{thebibliography}
\end{document}